\documentclass[10pt,prl,aps,showpacs,twocolumn,unsortedaddress]{revtex4-1}
\usepackage[caption=false]{subfig}
\usepackage{amssymb}
\usepackage{amsmath}
\usepackage{commath}
\usepackage{graphicx,bm}
\usepackage{verbatim}
\usepackage[usenames]{color}
\usepackage{pdfpages} 
\usepackage{hyperref}

\renewcommand{\vec}[1]{\mathbf{#1}}

\newcommand{\expp}[1]{\text{exp}\left(#1\right)}
\def\be#1\ee{\begin{equation}#1\end{equation}}
\def\ba#1\ea{\begin{align}#1\end{align}}
\newcommand{\eqqref}[1]{\text{Eq.}\,\eqref{#1}}
\renewcommand{\figref}[1]{\text{Fig.}\,\ref{#1}}

\begin{document}
\title{Interaction Effects on Number Fluctuations in a Bose-Einstein Condensates of Light}
\author{E.C.I. van der Wurff}
\email{e.c.i.vanderwurff@students.uu.nl}
\author{A.-W. de Leeuw}
\author{R.A. Duine}
\author{H.T.C. Stoof}
\affiliation{Institute for Theoretical Physics and Center for Extreme Matter and Emergent Phenomena, Utrecht University, Leuvenlaan 4, 3584 CE Utrecht, The Netherlands}
\date{\today}
\begin{abstract}
We investigate the effect of interactions on condensate-number fluctuations in Bose-Einstein condensates. For a contact interaction we variationally obtain the equilibrium probability distribution for the number of particles in the condensate. To facilitate comparison with experiment, we also calculate the zero-time delay autocorrelation function $g^{(2)}(0)$ for different strengths of the interaction. Finally, we focus on the case of a condensate of photons and discuss possible mechanisms for the interaction.
\end{abstract}



\maketitle
\textit{Introduction.---} 
Fluctuations are ubiquitous in physics: from the primordial quantum fluctuations in the early universe that reveal themselves as fluctuations in the cosmic microwave background, to current fluctuations in every-day conductors. For large voltages, the latter fluctuations give rise to shot noise, that is due to the discrete nature of charge \cite{Beenakker}. As a consequence, shot noise can be used to determine the quanta of the electric charge of the current carriers in conducting materials \cite{Schottky}. Indeed, it has been used to characterize the nature of Cooper pairs in superconductors \cite{Cooper} and the fractional charge of the quasiparticles of the quantum Hall effect \cite{Frac}. For low voltages, the noise in the current is thermal and is called Johnson-Nyquist noise \cite{Nyquist,Johnson}. Contrary to shot noise, thermal noise is always present in electrical circuits, even if no externally applied voltage is present, since it is due to thermal agitation of charge carriers, that leads to fluctuating electromotive forces in the material.

Theoretically, fluctuations in equilibrium are described by the fluctuation-dissipation
theorem, as formulated by Nyquist in 1928 and proven decades later \cite{CalWel}. This theorem relates the response of a system to an external perturbation to the fluctuations in the system in the absence of that perturbation. Given a certain fluctuation spectrum we can reconstruct the response of the system. Therefore, this theorem is very powerful, as was fervently argued by the Japanese physicist Kubo \cite{Kubo}.

Having stressed the importance of fluctuations in physics and the information they contain, we now zoom in on condensate-number fluctuations as our main point of interest. Traditionally, weakly interacting Bose-Einstein condensates were first observed in dilute atomic vapors \cite{AtomicBEC1}. For these systems, it is very difficult to measure number fluctuations because typically number measurements are destructive. Therefore, theoretical work has focused more on density-density correlation functions \cite{AtomicBEC2,Noise}.

In recent years, Bose-Einstein condensates of quasiparticles have also been created, such as exciton-polariton condensates \cite{QuasiBEC1}, magnon condensates \cite{QuasiBEC2} and condensates of photons \cite{Weitz,Weitz3}. These condensates of quasiparticles are realized under different circumstances compared to the atomic condensates. For instance, these condensates are created at higher temperatures than the condensates of dilute atomic gases: from 19 K for the exciton-polariton condensate to room temperature for the photonic condensate. Additionally, the condensates of quasiparticles are not in true equilibrium, since the steady state is a dynamical balance between particle losses and particle gain by external pumping with a laser. Due to these differences, new experimental possibilities have opened up. For example, large number fluctuations of the order of the total particle number have been observed in a condensate of photons \cite{Weitz2}.

In this Letter we investigate number fluctuations in these new Bose-Einstein condensates. We start by introducing an effective contact interaction into the grand-canonical Hamiltonian of a Bose gas and derive an equilibrium probability distribution for the number of particles in the condensate. Subsequently, we investigate these distributions for different condensate fractions and interaction strengths. We also calculate the zero-time delay autocorrelation function $g^{(2)}(0)$ to quantify the number fluctuations and compare this to experiments. Finally, we focus on Bose-Einstein condensates of photons and discuss possible mechanisms for the interactions.

\textit{Interaction effects on number fluctuations.---} 
We consider a harmonically trapped Bose gas with a fixed number of particles. Because the condensates of quasiparticles are typically confined in one direction, we specialize to the case of two dimensions. However, the following treatment is completely general and can easily be generalized to higher or lower dimensions.

To investigate the number fluctuations, we first calculate the average number of particles $\langle N_0 \rangle$ in the condensate. Because the condensates of quasiparticles allow for a free exchange of bosons with an external medium we treat the system in the grand-canonical ensemble: the probability distribution $P(N_0)$ for the number of condensed particles is of the form $P(N_0)\propto \expp{-\beta \Omega(N_0)}$, with $\Omega(N_0)$ the grand potential of the gas of bosons.

To find the grand potential we use a variational wavefunction approach. We note that the bosons in the condensate typically interact with each other. A reasonable first approximation for the form of this interaction is a contact interaction, as essentially every interaction is renormalized to a contact interaction at long length and time scales, independent of the precise origin of the interactions. Therefore, we consider the following energy functional for the macroscopic wavefunction $\phi_0(\vec{x})$ of the Bose-Einstein condensate \cite{Stoof}
\ba \label{eq:enfunc}
\Omega[\phi_0(\vec{x})]&=\int \dif{\vec{x}} \bigg( \frac{\hbar^2}{2 m}\left |\nabla \phi_0(\vec{x})\right |^2 + V^{\text{ex}}(\vec{x}) \left| \phi_0(\vec{x})  \right|^2 \nonumber \\
&\phantom{=} \quad \qquad -\mu|\phi_0(\vec{x})|^2 + \frac{g}{2}\left|\phi_0(\vec{x})\right|^4  \bigg),
\ea
where $\vec{x}$ is the two-dimensional position, the first term represents the kinetic energy of the condensate, $V^{\text{ex}}(\vec{x}) = m\omega^2|\vec{x}|^2/2$ is the harmonic trapping potential, $\mu$ is the chemical potential for the particles and $g$ is the coupling constant of the effective pointlike interaction between the particles.

We use the Bogoliubov substitution $\phi_0(\vec{x})=\sqrt{N_0}\psi_q(\vec{x})$, with the normalized variational wavefunction $\psi_q(\vec{x})$, such that $\int \dif \vec{x}|\phi_0(\vec{x})|^2=N_0$. Subsequently, we minimize the energy as a function of the variational parameter $q$, which describes the width of the condensate. As an ansatz we take the variational wavefunction to be the Gaussian $\psi_q(\vec{x}) = (\sqrt{\pi}q)^{-1}\expp{-|\vec{x}|^2/2q^2}$. Substituting this into the energy given by \eqqref{eq:enfunc} and minimizing with respect to the variational parameter, we obtain
\be
q_{\text{min}} = \sqrt[4]{ \frac{2\pi\hbar^2 + m N_0 g}{2\pi \omega^2m^2}} = q_{\text{ho}}\sqrt[4]{1+\frac{\tilde{g}N_0}{2\pi}},
\ee
where we introduced the dimensionless coupling constant $\tilde{g} := m g/\hbar^2$ and the harmonic oscillator length $q_{\text{ho}}=\sqrt{\hbar/m \omega}$. Note that for a sufficiently small number of condensate particles $q_{\text{min}}$ reduces to $q_{\text{ho}}$. For a large number of condensate particles the Thomas-Fermi ansatz for the wavefunction is in principle more appropriate. However, it is well known from the atomic condensates \cite{GaussianAtomic} that even in this case the Gaussian approach is rather accurate.

We now substitute the minimal value for the variational parameter $q$ into the energy functional, yielding the probability distribution 
\be \label{eq:prob}
P(N_0) \propto \text{ }\expp{\beta N_0 \bigg(\mu - \hbar\omega \sqrt{ 1+ \frac{ \tilde{g} N_0 }{2\pi}}\bigg)},
\ee
where the normalization is $\int_0^{\infty}\dif N_0 P(N_0)=1$.

Experimentally, the relevant parameter is the condensate fraction $x:=\langle N_0 \rangle / \langle N \rangle$, with $N$ the total number of particles. Thus, to relate our results to the experiments we need a relation between $\langle N_0 \rangle $ and the average total number of particles. For temperatures $T$ below the critical temperature for Bose-Einstein condensation, the average number of particles in excited states can in a good approximation be determined from the ideal-gas result. We obtain
\be \label{eq:BEC}
\langle N_{\text{ex}}(T) \rangle=\int_0^\infty \frac{g(\epsilon)\dif \epsilon}{\expp{\epsilon/k_B T}-1}=\frac{N_s}{6}\bigg(\frac{\pi k_B T}{\hbar \omega}\bigg)^2,
\ee
where we used the density of states $g(\epsilon) = N_s\epsilon/(\hbar\omega)^2$ for a two-dimensional harmonic trapping potential \cite{Smith}. The integer $N_s$ denotes the number of spin components of the boson. The critical temperature $T_c$ is defined by $\langle N \rangle = \langle N_{\text{ex}}(T_c)\rangle $. With this criterion, we find
\be \label{eq:BEC2}
\langle N_0 \rangle = \frac{xN_s}{6(1 - x)}\bigg(\frac{\pi k_B T}{\hbar\omega}\bigg)^2.
\ee

\textit{Results.---} 
\begin{figure}[t]
\centering
\includegraphics[trim=0cm 0cm 1.2cm 0cm, clip=true,width=.45\textwidth]{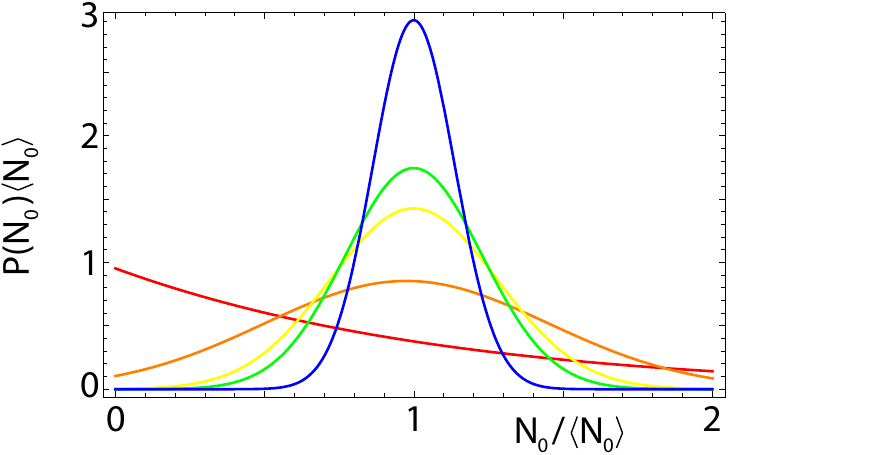}
\caption{(color online). Typical plot of the probability distribution for two-component bosons for a fixed interaction strength $\tilde{g} = 5\cdot 10^{-6}$ and different condensate fractions $x_{\text{red}}$ = 0.04, $x_{\text{orange}}= 0.28$, $x_{\text{yellow}}=0.40$, $x_{\text{green}}=0.45$ and $x_{\text{blue}}=0.58$.}
\label{fig:prob}
\end{figure}
Given an interaction strength $\tilde{g}$, we use the normalized probability distribution in \eqqref{eq:prob} to calculate the chemical potential as a function of $\langle N_0 \rangle$, i.e. $\mu=\mu(\langle N_0 \rangle)$. Given a condensate fraction $x$, we then use \eqqref{eq:BEC2} to calculate $\langle N_0 \rangle$ and the corresponding $\mu$. As an example we take $N_s=2$, which is approriate for the Bose-Einstein condensate of photons \cite{Weitz,Weitz3,Weitz2}. Finally, we use the obtained chemical potential to plot the probability distribution at fixed $x$ and $\tilde{g}$. Typical plots of the probability distribution for different condensate factions are displayed in \figref{fig:prob}. Clearly, we have exponential behavior due to a Poissonian process for small condensate fractions and Gaussian behavior for larger condensate fractions. Physically, this shows that the effect of repulsive interactions is to reduce number fluctuations, as the interactions give fluctuations an energy penalty. Increasing the interaction strength yields Gaussian behavior for even smaller condensate fractions. These Gaussians are also more strongly peaked around $\langle N_0 \rangle$ for higher interaction strengths, which is expected since stronger interactions between the bosons leads to the surpression of fluctuations.

\begin{figure}[t]
\centering
\includegraphics[trim=0cm 0cm 0cm 0cm, clip=true,width=.45\textwidth]{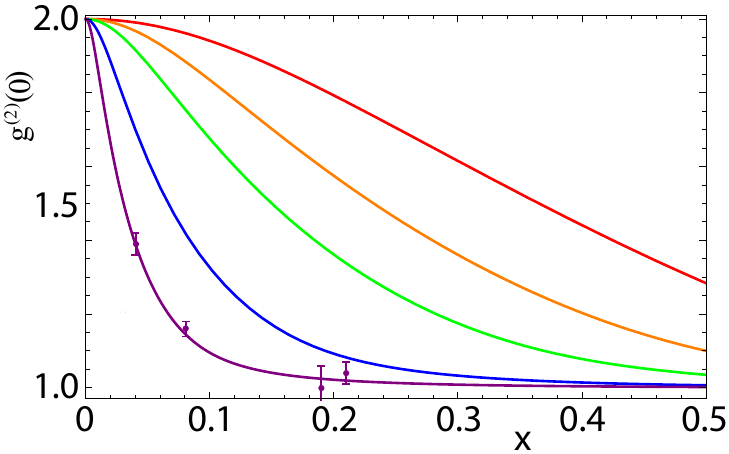}
\caption{(color online). Plot of the zero-time delay autocorrelation function $g^{(2)}(0)$ against the condensate fraction $x$ for $\omega = 8 \pi\cdot10^{10}$ Hz and $T=300$ K. The different curves correspond to different interaction strengths: $\tilde{g}_{\text{red}}= 5\cdot 10^{-7}$, $\tilde{g}_{\text{orange}}=2\cdot 10^{-6}$, $\tilde{g}_{\text{green}}=5\cdot 10^{-6}$, $\tilde{g}_{\text{blue}}=3\cdot 10^{-5}$, $\tilde{g}_{\text{purple}}=2\cdot 10^{-4}$. The purple curve is fit to the included experimental points from Klaers \emph{et al.} \cite{Weitz2}.}
\label{fig:auto}
\end{figure}

Next, we obtain the second moment $\langle N_0^2 \rangle$ from the probability distribution $P(N_0)$. This gives us all the information needed to quantify the number fluctuations of the condensate in terms of the zero-time delay autocorrelation function $g^{(2)}(0)$, which is defined as
\be
g^{(2)}(0) :=\frac{\langle N_0^2\rangle}{\langle N_0 \rangle^2}. 
\ee
A plot of this quantity against the condensate fraction is displayed in \figref{fig:auto} for different interaction strengths $\tilde{g}$. We note that bunching of bosons takes place for all interactions at small condensate fractions. For larger condensate fractions $g^{(2)}(0)\rightarrow 1$. The interpretation is as follows. Suppose we fix the condensate fraction $x$. At small interactions the quartic term in the energy in $\eqqref{eq:enfunc}$ is small and the minima of the energy are small and broad, yielding large number fluctuations. If we increase the interaction, the minima become deeper and more narrow, effectively reducing the fluctuations. The same reasoning holds for a fixed interaction strength and increasing condensate fractions, as we can also see in \figref{fig:prob}.

\textit{Discussion.---}
The results in the previous sections were quite generic for a two-dimensional, harmonically trapped gas of bosons with two possible polarizations. In fact, measurements of $g^{(2)}(0)$ have been performed recently \cite{Weitz2} in a Bose-Einstein condensate of photons, enabling us to compare our theory with experiments. In this experiment photons are confined in a dye-filled cavity, providing a harmonic potential and giving the photons an effective mass $m$ by fixing their longitudinal momentum $k_z$ \cite{Weitz}. The photons thermalize to the temperature of the dye solution by scattering of the dye molecules. Additionally, photon losses from the cavity are compensated by external pumping, yielding a constant average number of photons.

In \figref{fig:auto} we plot the data of Klaers \emph{et al.} for the situation that the photon interactions are experimentally known \cite{Weitz}. This data set is closest to our theoretical curve with $\tilde{g}_{\text{purple}}=2\cdot 10^{-4}$. By measuring the size of the condensate for different condensate fractions, it was experimentally found that $\tilde{g}=(7\pm3)\cdot10^{-4}$, which is reasonably close to our result. It must be noted that the experimental conditions of the included data points in \figref{fig:auto} were not identical to those in the measurement of the interaction strength. This is important, because deviations in the value of the trapping frequency $\omega$ can change the outcome of $\tilde{g}$ significantly.

Klaers \emph{et al.} have also studied the dependence of number fluctuations on the dye molecule density $n_{\text{mol}}$ and detuning $\delta$, which is roughly the difference between the cavity frequency and a dye specific frequency related to the effective absorption threshold of the dye molecules. Within our theory, the dependence of number fluctuations on these parameters can be incorporated via their influence on the interactions. Unfortunately, the dependence of the interactions on detuning and density of dye molecules is not experimentally known. Therefore, it would be useful to perform systematic measurements of $\tilde{g}$ for different detunings and molecule concentrations, as is also proposed in Ref.\,\cite{Engeland}. With this information, we would be able to directly compare all experimental results with our theoretical predictions for the number fluctuations.

The question remains what mechanism can cause an interaction that depends on both $n_{\text{mol}}$ and the detuning $\delta$. In fact, we conclude from the experimental data in Ref.\,\cite{Weitz2} that the interaction behaves counter-intuitively: it decreases both for an increasing molecule density and for a decreasing detuning. It has been suggested that the interaction is caused by slight changes in the refractive index of the solution as a function of either temperature or the intensity of the photons \cite{Private,Engeland}. The former phenomenon is known as thermal lensing, whereas the latter is known as a Kerr-type nonlinearity. We shall now focus on both effects seperately and estimate the strength of the resulting interactions.

Thermal lensing is the phenomenon that the index of refraction $n$ depends on the temperature of the medium. In the experiment of interest to us, non-radiative decay of the dye molecules, local fluctuations in the photon number and the external pumping with a laser lead to temperature fluctuations around the average temperature $T_0$. For a homogenous temperature distribution this implies, to lowest order that $n(T) = n(T_0) + \alpha (T-T_0)$. As the photon energy depends on the index of refraction, these temperature fluctuations couple to the photons. This leads to a photon-photon interaction as displayed in the Feynman diagram in \figref{fig:temp}. By assuming that the temperature fluctuations behave diffusively, we derive in the supplemental material that the interaction strength due to this effect is given by
\begin{figure}[t]
\centering
\includegraphics[trim=5.5cm 18.5cm 8.7cm 4.3cm, clip=true,width=.29\textwidth]{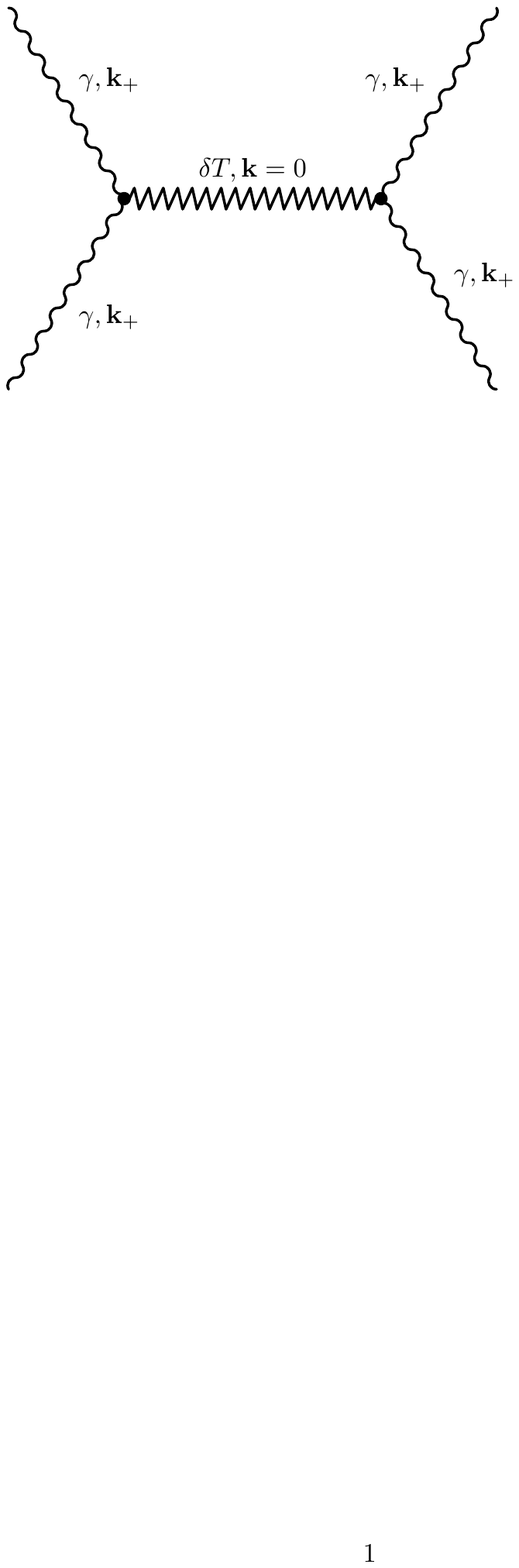}
\caption{Feynman diagram for the photon-photon interaction due to the diffusion of temperature fluctuations. The photons $\gamma$ are considered to be part of the condensate and are thus at zero frequency and at momentum $\vec{k_+}=(0,0,k_z)$, as their z-component momentum is fixed and $k_x=k_y=0$ for the condensate of the homogeneous photon gas.}
\label{fig:temp}
\end{figure}
\be
\tilde{g} = \frac{4m^3 c^4 \alpha^2 T_0 }{3D_0 \hbar^2 n^6(T_0) c_p},
\ee
where $D_0 = 7\pi/k_z$ is the length scale associated to the fixed longitudinal momentum $k_z$ of the photons and $c_p$ is the heat capacity of the solution. Note that this interaction has no explicit dependence on the detuning $\delta$, or on the concentration of dye molecules $n_{\text{mol}}$, as it is fully determined by the properties of the solution. Both the temperature dependence of the single-photon energy and the heat capacity might depend on the number of dissolved dye molecules, but as the experiments by Klaers \emph{et al.} are performed for small concentrations of dissolved Rhodamine 6G, we expect at least this latter effect to be small. Neglecting the former effect and using typical numerical values for liquid methanol with 1 mmol Rhodamine 6G dissolved \cite{HandBook,Turkish}, we obtain an estimate for the interaction strength of $\tilde{g} \sim 10^{-9}$. This is several orders of magnitude below the only experimental result $\tilde{g} \sim 10^{-4}$.

\begin{figure}[t]
\centering
\includegraphics[trim=5.5cm 17.6cm 9cm 4.4cm, clip=true,width=.25\textwidth]{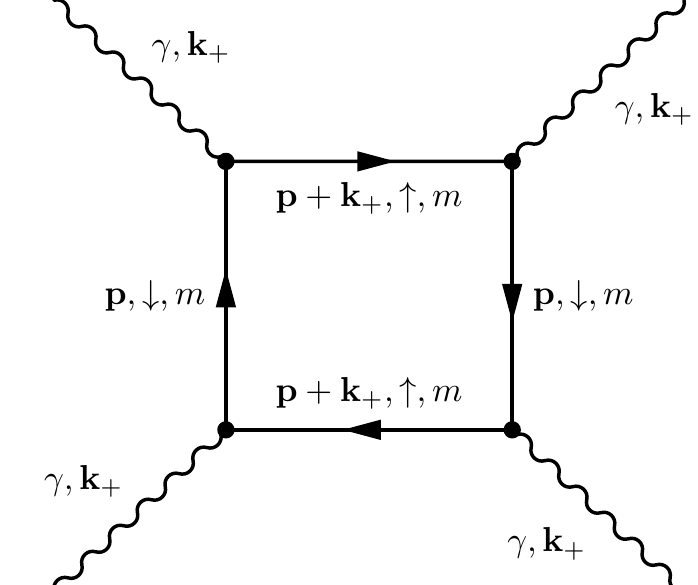}
\caption{Feynman diagram for the photon-photon interaction mediated by the dye molecules. Again we take the photons to be in the condensate: $\vec{k_+}=(0,0,k_z)$ and the frequency is zero. The molecule forms a closed loop of ground ($\downarrow$) and excited ($\uparrow$) states, with momentum $\vec{p}$ and Matsubara frequency $\omega_m$.}
\label{fig:box}
\end{figure}

The other possible photon-photon interaction is due to the Kerr effect: the index of refraction is changed by photon-photon scattering mediated by the dye molecules. It has been investigated in a somewhat different context and manner in Refs.\,\cite{PhotonPhoton,Engeland}. We choose to describe and calculate the Kerr effect by a Feynman diagram in the form of a box, as is shown in \figref{fig:box}.

In our earlier work in Ref.\,\cite{AW}, we adopted a simplified description of the complex rovibrational structure of the dye molecules by describing them as an effective two-level system and by giving the molecules an effective mass. Following this treatment, it turns out that the box diagram contains a divergence of the form $(\mu-\delta)^{-1}$. This is similar to the Feshbach resonances known from cold-atom physics \cite{Stoof,Feshbach}. To get around this non-physical divergence, we introduce a finite decay rate $\Gamma$ for the excited molecules. We show in the supplemental material that this leads to
\be \label{eq:TT}
\tilde{g}(\mu) = \frac{m g_{\text{mol}}^4\beta n_{\text{mol}}}{\hbar^4\Gamma^2 D_0} f(\mu-\delta),
\ee
where $g_{\text{mol}}$ is the coupling strength of the photons to the molecules and $f(\mu-\delta)$ is a smooth dimensionless function peaked around zero. Similar to the procedure followed in Ref.\,\cite{AW} we calculate the self-energy of the photons and by fitting to the seperately measured experimental absorption spectrum of the used dye we obtain $g_{\text{mol}}$, $\Gamma$ and $\delta$. Due to the introduction of the finite lifetime $\Gamma$, $\tilde{g}$ is no longer divergent, but simply peaked around the detuning $\delta$.

Subsequently, we have to solve $\tilde{g}(\mu)$ self-consistently with the Gross-Pitaevskii equation. Considering the center of the trap, i.e., $V^{\text{ex}}=0$, this amounts to solving $\tilde{g}(\mu) = (m/\hbar^2n_{\text{ph}})\mu$ for $\mu$, with $n_{\text{ph}}$ the photon density. Having solved this equation, we substitute the found chemical potential $\mu$ back into \eqqref{eq:TT} and obtain the self-consistent interaction strength. For typical experimental parameters we obtain $\tilde{g}\sim10^{-8}$, which is also small compared to the experimental value. However, the magnitude of $\tilde{g}$ is rather uncertain due to the simplification of the rovibrational energy spectrum of the dye molecules to a two-level system. Interestingly, we see from \eqqref{eq:TT} that this interaction depends both on the detuning $\delta$ and the density of molecules $n_{\text{mol}}$. We show in the supplemental material that a self-consistent solution can give rise to an interaction that decreases both for decreasing $\delta$ and increasing $n_{\text{mol}}$, which is precisely the counter-intuitive behavior the experimental results exhibit.

In conclusion, we have calculated the effect of self-interactions on number fluctuations in Bose-Einstein condensates. We have shown that the number fluctuations increase for decreasing interaction strengths. Furthermore, we have compared our results with a Bose-Einstein condensate of photons. We found rather good agreement with the experimental curve for which both the number fluctuations and $\tilde{g}$ are known. Subsequently, we discussed possible mechanisms for the photon-photon interaction. However, systematic measurements of the interaction strength are necessary to understand the true nature of the interaction. If the interaction is indeed a contact interaction at long wavelengths, then this would imply that the photon condensate is also a superfluid.

It is a pleasure to thank Dries van Oosten, Jan Klaers and Martin Weitz for useful discussions and the latter two also for providing experimental data. This work is supported by the Stichting voor Fundamenteel Onderzoek der Materie (FOM) and is part of the D-ITP consortium, a program of the Netherlands Organisation for Scientific Research (NWO) that is funded by the Dutch Ministry of Education, Culture and Science (OCW).

\onecolumngrid
\newpage
\newpage


\section*{Supplementary material}

# Supplemental Material for "Interaction Effects on Number Fluctuations in a Bose-Einstein Condensate of Light"

E.C.I. van der Wurff,* A.-W. de Leeuw, R. A. Duine, and H. T. C. Stoof

*Institute for Theoretical Physics and Center for Extreme Matter and Emergent Phenomena,*
*Utrecht University, Leuvenlaan 4, 3584 CE Utrecht, The Netherlands*

(Dated: April 9, 2014)

In this supplemental material we provide detailed derivations of and discussions on possible photon-photon interactions in the condensate of photons. First, we consider thermal lensing and derive the corresponding interaction strength. Subsequently, we calculate the interaction strength for the Kerr effect. We discuss how the latter interaction effect depends the detuning $\delta$ and density of molecules $n_{\rm mol}$.

## CONTENTS

## I. INTRODUCTION

The goal of this supplemental material is to derive the equations for the interaction strengths that we used in the Letter. We start with the phenomenon of thermal lensing by considering an action that includes temperature fluctuations which behave diffusively. Integrating out these fluctuations, we arrive at a four-point vertex for the photons and calculate the resulting interaction strength for the photons in the condensate. Next, we investigate the Kerr effect by considering an action that incorporates the coupling between the photons and the dye molecules. By integrating out the molecular fields, we arrive at an effective action for the photons, including a self-energy and a four-point interaction vertex for the photons. Via an explicit calculation of the self-energy of the photons, we are able to fit the coupling strength between atoms and molecules, the lifetime of the excited state and the detuning to the experimentally known absorption cross section. Hereafter, we give an explicit expression for the box diagram and we obtain numerical values for the interaction strength by using these particular values of the parameters. We end by discussing how this latter interaction depends on the detuning $\delta$ and density of molecules $n_{\rm mol}$.

## II. THERMAL LENSING

Thermal lensing is the phenomenon that an index of refraction is a function of the temperature of the system. In the case of a homogenous temperature distribution this implies to lowest order that $n(T) = n(T_0) + \alpha(T - T_0)$. However, if the temperature fluctuates, we have $T(\mathbf{x},t) = T_0 + \delta T(\mathbf{x},t)$, where $T_0 := \langle T(\mathbf{x},t) \rangle$ is the average temperature of the system. We recall that the photons in the photonic condensate have a fixed longitudinal direction $k_z$, such that their

---

*e.c.i.vanderwurff@students.uu.nl

energy is given by [1–3]

$$E = \frac{\hbar c}{n}\sqrt{k_z^2 + k_r^2}$$
$$\approx \frac{mc^2}{n^2} + \frac{\hbar^2 k_r^2}{2m} + \frac{1}{2}m\omega^2|\mathbf{r}|^2, \tag{1}$$

with $m := k_z\hbar n/c$, $k_r$ the two-dimensional transversal momentum, $\mathbf{r} = (x,y)$ a radial vector and $\omega := c\sqrt{2/D_0 R n^2}$ the harmonic trapping frequency with $R$ the curvature of the mirrors. We now consider the following action in imaginary time for the photon field $\phi(\mathbf{x},\tau)$ and temperature fluctuation field $\delta T(\mathbf{x},\tau)$

$$S = \int d\mathbf{x} \int d\tau \left[\phi^*(\mathbf{x},\tau)\left(\hbar\frac{\partial}{\partial\tau} - \frac{\hbar^2\nabla_\mathbf{r}^2}{2m} - \frac{mc^2}{n^2} + \frac{1}{2}m\omega^2|\mathbf{r}|^2 - \mu\right)\phi(\mathbf{x},\tau) + \frac{\delta T(\mathbf{x},\tau)}{2T_0}\left(c_p + \frac{\kappa\nabla^2}{i\partial_\tau}\right)\delta T(\mathbf{x},\tau)\right], \tag{2}$$

where $\mathbf{x} = (x,y,z)$ is a three-dimensional vector, $\nabla_\mathbf{r}^2$ denotes that we only consider motion in the transversal direction, $c_p$ is the heat capacity and $\kappa$ the thermal conductivity. The part which is quadratic in the temperature fluctuations is constructed such that it has the correct diffusive pole and diffusion propagator.

We now write for the index of refraction $n(T) = n(T_0) + \alpha\delta T(\mathbf{x},\tau)$ to include thermal lensing, whilst keeping $k_z$ fixed. Substituting this into the action and expanding for small temperature fluctuations, we obtain

$$S = \int d\mathbf{x}\int d\tau \phi^*(\mathbf{x},\tau)\left(\hbar\frac{\partial}{\partial\tau} - \frac{\hbar^2\nabla_\mathbf{r}^2}{2m} - \frac{mc^2}{n^2(T_0)} + \frac{1}{2}m\omega^2|\mathbf{r}|^2 - \mu\right)\phi(\mathbf{x},\tau)$$
$$+ \int d\mathbf{x}\int d\tau \frac{\delta T(\mathbf{x},\tau)}{2T_0}\left(c_p + \frac{\kappa\nabla^2}{i\partial_\tau}\right)\delta T(\mathbf{x},\tau) + \frac{2mc^2\alpha}{n^3(T_0)}\int d\mathbf{x}\int d\tau \delta T(\mathbf{x},\tau)\phi^*(\mathbf{x},\tau)\phi(\mathbf{x},\tau). \tag{3}$$

We shift away the constant offset in the energy by setting $\mu' = \mu + mc^2/n^2(T_0)$. Furthermore we note that the field $\phi(\mathbf{x},\tau)$ has an equation of motion in the longitudinal direction which decouples, such that we may write $\phi(\mathbf{x},\tau) = \phi_{\text{long}}(z)\phi_{\text{trans}}(\mathbf{r},\tau)$ with $\mathbf{r} \in \mathbb{R}^2$. The longitudinal part of the photon field is a standing wave and demanding that it vanishes at the boundaries of the cavity we obtain the normalized solution

$$\phi_{\text{long}}(z) = \sqrt{2/D_0}\sin(q\pi z/D_0), \tag{4}$$

with $D_0$ the fixed length of the cavity in the longitudinal direction and $q = 7,8$ in the experiments of interest to us [1–3]. As $\int dz|\phi_{\text{long}}(z)|^2 = 1$, we find

$$S = \int d\mathbf{r}\int d\tau \phi^*_{\text{trans}}(\mathbf{r},\tau)\left(\hbar\frac{\partial}{\partial\tau} - \frac{\hbar^2\nabla_\mathbf{r}^2}{2m} + \frac{1}{2}m\omega^2|\mathbf{r}|^2 - \mu'\right)\phi_{\text{trans}}(\mathbf{r},\tau)$$
$$+ \int d\mathbf{x}\int d\tau \frac{\delta T(\mathbf{x},\tau)}{2T_0}\left(c_p + \frac{\kappa\nabla^2}{i\partial_\tau}\right)\delta T(\mathbf{x},\tau) + \frac{2mc^2\alpha}{n^3(T_0)}\int d\mathbf{x}\int d\tau \delta T(\mathbf{x},\tau)\phi^*(\mathbf{x},\tau)\phi(\mathbf{x},\tau). \tag{5}$$

We define the Fourier transforms as

$$\begin{cases} \phi(\mathbf{x},\tau) = (\hbar\beta V)^{-1/2}\sum_{\mathbf{p},n} a_{\mathbf{p},n}e^{i(\mathbf{p}\cdot\mathbf{x}-\omega_n\tau)}, \\ \phi_{\text{trans}}(\mathbf{r},\tau) = (\hbar\beta A)^{-1/2}\sum_{\mathbf{k},n} a_{\mathbf{k},n}e^{i(\mathbf{k}\cdot\mathbf{r}-\omega_n\tau)}, \\ \delta T(\mathbf{x},\tau) = (\hbar\beta V)^{-1/2}\sum_{\mathbf{p},n}\delta T_{\mathbf{p},n}e^{i(\mathbf{p}\cdot\mathbf{x}-\omega_n\tau)}, \end{cases} \tag{6}$$

with the volume $V$ and the area $A := V/D_0$. We use the convention that a sum over $\mathbf{k}$ is over all vectors with a fixed z-component $k_z$ and sums over $\mathbf{p}$ are ordinary three-dimensional sums. Substituting these Fourier transforms into the action, we arrive at

$$S = \sum_{\mathbf{k},n} a^*_{\mathbf{k},n}(-\hbar G_\gamma^{-1}(\mathbf{k},i\omega_n))a_{\mathbf{k},n} + \sum_{\mathbf{p},n}\delta T^*_{\mathbf{p},n}(-\hbar G_T^{-1}(\mathbf{p},i\omega_n))\delta T_{\mathbf{p},n}$$
$$+ \frac{1}{\sqrt{\hbar\beta V}}\left(\frac{mc^2\alpha}{n^3(T_0)}\right)\sum_{\mathbf{k},\mathbf{p},n,m}\left(a^*_{\mathbf{k}+\mathbf{p},m+n}a_{\mathbf{k},n}\delta T_{\mathbf{p},m} + a^*_{\mathbf{k},n}a_{\mathbf{k}+\mathbf{p},m+n}\delta T^*_{\mathbf{p},m}\right), \tag{7}$$

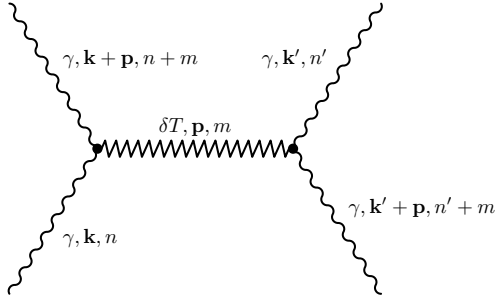

FIG. 1. Feynman diagram for the photon-photon interaction due to the diffusion of temperature fluctuations.

where we used $\delta T^*_{\mathbf{p},n} = \delta T_{-\mathbf{p},-n}$. In the process we defined the inverse propagator in Fourier space for the temperature fluctuations

$$-\hbar G_T^{-1}(\mathbf{p},\omega_n) = \frac{1}{T_0}\left(c_p - \frac{\kappa|\mathbf{p}|^2}{\omega_n}\right). \tag{8}$$

Subsequently, we complete the square, perform the path integral over $\delta T_{\mathbf{k},n}$ and re-exponentiate to arrive at the effective action

$$S = \sum_{\mathbf{k},n} a^*_{\mathbf{k},n}(-\hbar G_\gamma^{-1}(\mathbf{k},i\omega_n))a_{\mathbf{k},n} - \hbar \mathrm{Tr}\log(-G_T^{-1}(\mathbf{p},\omega_n))$$
$$-\frac{1}{\hbar^2 \beta V}\left(\frac{mc^2\alpha}{n^3(T_0)}\right)^2 \sum_{\mathbf{k},\mathbf{k}',\mathbf{p}}\sum_{n,n',m} a^*_{\mathbf{k}+\mathbf{p},n+m}a_{\mathbf{k},n}a^*_{\mathbf{k}',n'}a_{\mathbf{k}'+\mathbf{p},n'+m}G_T(\mathbf{p},\omega_m). \tag{9}$$

The last term is depicted as a Feynman diagram in Fig. 1. From the effective action we read off the photon-photon four-point vertex

$$\Gamma^{(4)}(\mathbf{p},m) = -\frac{2}{\hbar}G_T(\mathbf{p},\omega_m)\left(\frac{mc^2\alpha}{n^3(T_0)}\right)^2. \tag{10}$$

As the photon gas is confined to two dimensions we must scale $\Gamma^{(4)}(\mathbf{p},m) \to \left(\int_0^{D_0}|\phi_{\mathrm{long}}(z)|^4\,\mathrm{d}z\right)\Gamma^{(4)}(\mathbf{p},m) = 2\Gamma^{(4)}(\mathbf{p},m)/3D_0$ to obtain the effective coupling constant $g$ of the photons in the condensate. Taking the photons in the condensate, i.e., with Matsubara frequency zero and only a non-zero momentum of $k_z$ in the z-direction, we obtain

$$\tilde{g} = \frac{2\Gamma^{(4)}(\mathbf{0},0)m}{3\hbar^2 D_0} = \frac{4m^3 c^4\alpha^2 T_0}{3D_0\hbar^2 n^6(T_0)c_p}. \tag{11}$$

Typical values for the experimental parameters are: $m = 6.7\cdot 10^{-36}$ kg, $T_0 = 300$ K, $D_0 = 1.46\cdot 10^{-6}$ m [1], $n = 1.34$, $\alpha = -5\cdot 10^{-4}$ K$^{-1}$ [4], $c_p = \tilde{c}_p/V_m$ with $\tilde{c}_p = 79.5$ J mol$^{-1}$ K$^{-1}$ and $V_m = 40.0\cdot 10^{-6}$ m$^3$ mol$^{-1}$ [5]. These values yield an estimate for the interaction strength of $\tilde{g} \sim 10^{-9}$.

### III. THE KERR EFFECT

We now neglect the temperature depence of the index of refraction and focus on the Kerr effect. We therefore consider an Euclidean action which includes the interactions between the photons and the molecules [6]

$$S = \sum_{\mathbf{k},n}a^*_{\mathbf{k},n}(-i\hbar\omega_n + \epsilon_\gamma(\mathbf{k}) - \mu)a_{\mathbf{k},n} + \sum_{\mathbf{p},\rho,n}b^*_{\mathbf{p},\rho,n}(-i\hbar\omega_n + \epsilon(\mathbf{p}) - \mu_\rho + K_\rho)b_{\mathbf{p},\rho,n}$$
$$+\frac{g_{\mathrm{mol}}}{\sqrt{\hbar\beta V}}\sum_{\mathbf{k},\mathbf{p},n,n'}\left(a_{\mathbf{k},n}b_{\mathbf{p},\downarrow,n'}b^*_{\mathbf{p}+\mathbf{k},\uparrow,n+n'} + a^*_{\mathbf{k},n}b_{\mathbf{p}+\mathbf{k},\uparrow,n+n'}b^*_{\mathbf{p},\downarrow,n'}\right)$$
$$:= S_\gamma + S_0 + S_{\mathrm{int}}, \tag{12}$$

where $V$ is the three-dimensional volume of the system, $\beta := 1/k_B T$ is the inverse thermal energy, $g_{\text{mol}}$ is the coupling constant between the atoms and the molecules, $\epsilon(\mathbf{p}) = \hbar^2|\mathbf{p}|^2/2M$ is the dispersion relation for the molecules with mass $M$ and $\epsilon_\gamma(\mathbf{k}) = \hbar c_{\text{med}}\sqrt{k_x^2 + k_y^2 + k_z^2}$ the dispersion relation for the photons, in which $k_z$ is the fixed longitudinal momentum and $c_{\text{med}}$ the speed of light in the medium. In the action, we introduced the photon field amplitude $a_{\mathbf{k},n}$ and molecule field amplitude $b_{\mathbf{p},\rho,n}$. The photon fields are bosonic and we model the molecules as fermions, although this is not important since in the end we always consider the classical limit for the dye molecules. Additionally, we model the molecules as a two-level system consisting of an excited state ($\uparrow$) and ground state ($\downarrow$). Furthermore, the associated energies of the states are given by $K_\uparrow = \Delta$ and $K_\downarrow = 0$. Again we use the convention that a sum over $\mathbf{p}$ is three-dimensional, whereas a sum over $\mathbf{k}$ is a two-dimensional sum over a three-dimensional vector with a fixed z-component. Finally, the last two terms describe the absorption and emission of a photon.

From the action we read off the propagators of the non-interacting theory in Fourier space

$$\begin{cases} G_\gamma(\mathbf{k}, i\omega_n) = -\hbar(-i\hbar\omega_n + \epsilon_\gamma(\mathbf{k}) - \mu)^{-1}, \\ G_\rho(\mathbf{k}, i\omega_n) = -\hbar(-i\hbar\omega_n + \epsilon(\mathbf{k}) - \mu_\rho + K_\rho)^{-1}, \end{cases} \quad (13)$$

where $\omega_n$ denote the appropriate Matsubara frequencies. By using the action, we write down the partition function $Z$ of the theory as a path integral over the photonic and molecular fields. Subsequently, we perform perturbation theory in the interaction parameter $g_{\text{mol}}$ to integrate out the molecules [7], i.e.,

$$Z = \int d[a^*]\,d[a]\,d[b_\downarrow^*]\,d[b_\downarrow]\,d[b_\uparrow^*]\,d[b_\uparrow] \exp\left(-\frac{1}{\hbar}(S_\gamma + S_0 + S_{\text{int}})\right)$$
$$= Z_0 \int d[a^*]\,d[a]\exp\left(-\frac{1}{\hbar}S_\gamma\right)\left(1 + \frac{1}{2\hbar^2}\langle S_{\text{int}}^2\rangle_0 + \frac{1}{24\hbar^4}\langle S_{\text{int}}^4\rangle_0 + \mathcal{O}(g_{\text{mol}}^6)\right), \quad (14)$$

where we defined $\langle \ldots \rangle_0 := Z_0^{-1} \int d[b_\downarrow^*]\,d[b_\downarrow]\,d[b_\uparrow^*]\,d[b_\uparrow](\ldots)\exp(-S_0/\hbar)$, $Z_0 := \int d[b_\downarrow^*]\,d[b_\downarrow]\,d[b_\uparrow^*]\,d[b_\uparrow]\exp(-S_0/\hbar)$ and used the fact that $\langle S_{\text{int}}^m \rangle_0 = 0$ if $m$ is odd. By using Wick's theorem we find for the term at order $g_{\text{mol}}^2$

$$\langle S_{\text{int}}^2 \rangle = -\frac{2g_{\text{mol}}^2}{\hbar \beta V} \sum_{\mathbf{k},\mathbf{p},n,n'} a^*_{\mathbf{k},n} a_{\mathbf{k},n} G_\uparrow(\mathbf{p}+\mathbf{k}, i(\omega_n + \omega_{n'})) G_\downarrow(\mathbf{p}, i\omega_{n'}). \quad (15)$$

This term can be interpreted as a self-energy for the photons. A diagrammatic representation of this self-energy is depicted in Fig. 2. We do a similar computation for the term at order $g_{\text{mol}}^4$. Expanding out the interaction term yields

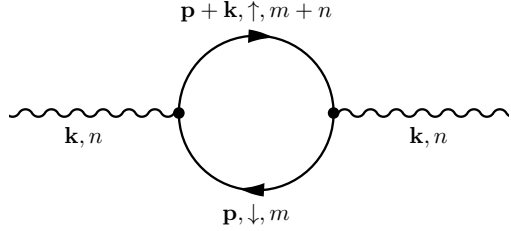

FIG. 2. The Feynman diagram corresponding to Eq. (15). This diagram represents the self-energy of the photons.

sixteen terms, of which ten do not contain the appropriate combination of fields to yield a non-zero result when we Wick contract them. The remaining six terms turn out to be identical, yielding

$$\langle S_{\text{int}}^4 \rangle_0 = \frac{12 g_{\text{mol}}^4}{(\hbar \beta V)^2} \sum_{\mathbf{p},\mathbf{k},n,n'} \sum_{\mathbf{k}',\mathbf{k}'',m,n''} a^*_{\mathbf{k},n} a_{\mathbf{k}',n'} a^*_{\mathbf{k}'',n''} a_{\mathbf{k}-\mathbf{k}'+\mathbf{k}'',n-n'+n''} G_\uparrow(\mathbf{p}+\mathbf{k}, i(\omega_m + \omega_n)) G_\downarrow(\mathbf{p}, i\omega_m)$$
$$\times G_\uparrow(\mathbf{p}+\mathbf{k}', i(\omega_m + \omega_{n'})) G_\downarrow(\mathbf{p}+\mathbf{k}'-\mathbf{k}'', i(\omega_m + \omega_{n'} - \omega_{n''}))$$
$$- \frac{12 g_{\text{mol}}^4}{(\hbar \beta V)^2} \left(\sum_{\mathbf{p},\mathbf{k},n,n'} a^*_{\mathbf{k},n} a_{\mathbf{k},n} G_\uparrow(\mathbf{p}+\mathbf{k}, i(\omega_n + \omega_{n'})) G_\downarrow(\mathbf{p}, i\omega_n)\right)^2. \quad (16)$$

The first term is diagrammatically displayed as a box diagram in Fig. 4. The second term can be represented by a

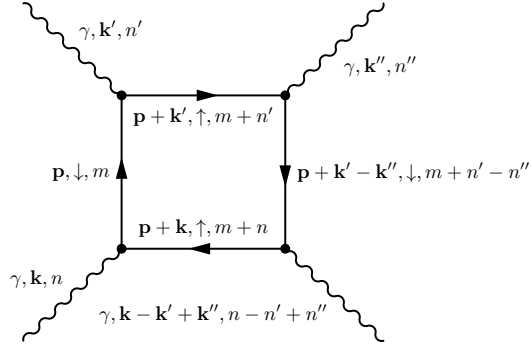

FIG. 3. The Feynman diagram corresponding to the first term in Eq. (16).

disconnected diagram: it is simply one half times the square of the self-energy diagram. All disconnected diagrams disappear automatically when we sum the connected diagrams into an exponent. Hence, the sum of the self-energy and the box diagram gives us the desired effective action

$$S^{\text{eff}} = \sum_{\mathbf{k},n} a^*_{\mathbf{k},n}(-i\hbar\omega_n + \epsilon_\gamma(\mathbf{k}) - \mu + \hbar\Sigma(\mathbf{k}, i\omega_n))a_{\mathbf{k},n}$$
$$+ \frac{1}{2\hbar\beta V} \sum_{\mathbf{k},\mathbf{k}',\mathbf{k}''} \sum_{n,n',n''} \Gamma^{(4)}(\mathbf{k}, \mathbf{k}', \mathbf{k}'', i\omega_n, i\omega_{n'}, i\omega_{n''}) a^*_{\mathbf{k},n} a_{\mathbf{k}',n'} a^*_{\mathbf{k}'',n''} a_{\mathbf{k}-\mathbf{k}'+\mathbf{k}'',n-n'+n''}, \quad (17)$$

where we defined the self-energy as

$$\hbar\Sigma(\mathbf{k}, i\omega_n) := \frac{g^2_{\text{mol}}}{\hbar^2\beta V} \sum_{\mathbf{p},m} G_\downarrow(\mathbf{p}, i\omega_m) G_\uparrow(\mathbf{p}+\mathbf{k}, i(\omega_n+\omega_m)), \quad (18)$$

and the photon-photon interaction vertex is given by

$$\Gamma^{(4)}(\mathbf{k}, \mathbf{k}', \mathbf{k}'', i\omega_n, i\omega_{n'}, i\omega_{n''}) := -\frac{g^4_{\text{mol}}}{\hbar^4\beta V} \sum_{\mathbf{p},m} G_\uparrow(\mathbf{p}+\mathbf{k}, i(\omega_m+\omega_n)) G_\downarrow(\mathbf{p}, i\omega_m)$$
$$\times G_\uparrow(\mathbf{p}+\mathbf{k}', i(\omega_m+\omega_{n'})) G_\downarrow(\mathbf{p}+\mathbf{k}'-\mathbf{k}'', i(\omega_m+\omega_{n'}-\omega_{n''})). \quad (19)$$

### A. Self-Energy

We are interested in calculating the self-energy given by Eq. (18). However, to avoid divergencies when calculating the photon-photon interaction we first introduce a finite lifetime for the excited molecular state. To do this we recall that the spectral function $\rho(\mathbf{k}, \omega)$ is defined as

$$\rho(\mathbf{k}, \omega) := -\frac{1}{\pi\hbar} \text{Im}\left[G^{(+)}(\mathbf{k}, \omega)\right], \quad (20)$$

where the retarded Green's function follows from a Wick rotation of the Green's function: $G^{(+)}(\mathbf{k}, \omega) = G(\mathbf{k}, i\omega_n \to \omega + i0)$. Given a spectral function, we calculate the corresponding Green's function by using the following relation

$$G_\uparrow(\mathbf{k}, i\omega_n) = \hbar \int_{-\infty}^{\infty} d\omega \frac{\rho_\uparrow(\mathbf{k}, \omega)}{i\omega_n - \omega}. \quad (21)$$

For a free theory the spectral function is just a delta function centered around the single-particle energy. We give the excited molecule a finite lifetime by broadening the spectral function to a Gaussian profile, i.e.,

$$\rho_\uparrow(\mathbf{k}, \omega) = \frac{1}{\sqrt{2\pi}\hbar\Gamma} \exp\left(-\frac{(\hbar\omega - \epsilon(\mathbf{k}) - \Delta + \mu_\uparrow)^2}{2(\hbar\Gamma)^2}\right), \quad (22)$$

5

such that the spectral function satisfies the frequency sum rule $\int_{-\infty}^{\infty} d\hbar\omega \rho_\uparrow(\mathbf{k},\omega) = 1$. Note that this spectral function is still centered around the single-particle energy of the excited molecule and that we have

$$\lim_{\Gamma \to 0} \rho_\uparrow(\mathbf{k},\omega) = \delta(\hbar\omega - \epsilon(\mathbf{k}) - \Delta + \mu_\uparrow). \tag{23}$$

We consider the molecules in the classical limit. Thus, by using the Maxwell-Boltzmann distribution $N_{\mathrm{MB}}(x) := \exp(-\beta x)$, the molecule density in the excited state $n_\uparrow$ is equal to

$$\begin{aligned}
n_\uparrow &\approx \frac{1}{V} \int_{-\infty}^{\infty} d\hbar\omega \sum_{\mathbf{k}} \rho_\uparrow(\mathbf{k},\omega) N_{\mathrm{MB}}(\hbar\omega) \\
&= \frac{1}{\Lambda^3} \exp\left(\beta\mu_\uparrow - \beta\Delta + \frac{1}{2}(\beta\hbar\Gamma)^2\right),
\end{aligned} \tag{24}$$

where the thermal de Broglie wavelength $\Lambda$ is defined as $\Lambda := \sqrt{2\pi\beta\hbar^2/M}$. As $\lim_{\omega\to-\infty} N_{\mathrm{FD}}(\omega) = 1$ and we integrate over $\omega$, the approximation in the calculation above is only valid when the spectral function is almost zero for negative $\omega$. As the spectral function is centered around $\hbar\omega = \epsilon(\mathbf{k}) + \Delta - \mu_\uparrow$, a reasonable restriction is $\Delta - \mu_\uparrow - 2\Gamma\hbar > 0$. This condition is fulfilled for the values of $\Delta$ and $\Gamma$ we use. We are thus allowed to take the limit $N_{\mathrm{FD}}(\omega) \to N_{\mathrm{MB}}(\omega)$ in the following, even if we integrate over $\omega$.

Note that in the ground state the molecules still have an infinite lifetime and therefore the density of molecules in the ground state is given by $n_\downarrow = \Lambda^{-3} \exp(\beta\mu_\downarrow)$. We express the chemical potential of the ground state in terms of $\Delta\mu := \mu_\uparrow - \mu_\downarrow$ and $n_{\mathrm{mol}} := n_\downarrow + n_\uparrow$ as

$$\exp(\beta\mu_\downarrow) = \frac{n_{\mathrm{mol}}\Lambda^3}{1 + \exp\left(\beta(\Delta\mu - \Delta) + (\hbar\beta\Gamma)^2/2\right)}, \tag{25}$$

which is a relation we will use later on. We now explicitly calculate the self-energy by starting from the definition provided in Eq. (18) and invoking Eqs. (21) and (22)

$$\begin{aligned}
\hbar\Sigma(\mathbf{k}, i\omega_n) &= \frac{g_{\mathrm{mol}}^2}{\hbar^2 \beta V} \sum_{\mathbf{p},m} G_\downarrow(\mathbf{p}, i\omega_m) G_\uparrow(\mathbf{p}+\mathbf{k}, i(\omega_m+\omega_n)) \\
&= \frac{g_{\mathrm{mol}}^2}{\hbar\beta V} \sum_{\mathbf{p},m} \int_{-\infty}^{\infty} d\hbar\omega' \left(\frac{\rho_\uparrow(\mathbf{p}+\mathbf{k},\omega')}{i\hbar(\omega_n+\omega_m) - \hbar\omega'}\right) \left(\frac{-\hbar}{-i\hbar\omega_m + \epsilon(\mathbf{p}) - \mu_\downarrow}\right) \\
&= \frac{g_{\mathrm{mol}}^2}{V} \sum_{\mathbf{p}} \int_{-\infty}^{\infty} d\hbar\omega' \left(\frac{\rho_\uparrow(\mathbf{p}+\mathbf{k},\omega')}{-i\hbar\omega_n + \hbar\omega' - \epsilon(\mathbf{p}) + \mu_\downarrow}\right) \left(N_{\mathrm{FD}}(\hbar\omega') - N_{\mathrm{FD}}(\epsilon(\mathbf{p}) - \mu_\downarrow)\right),
\end{aligned} \tag{26}$$

where we performed the Matsubara summation and introduced the Fermi-Dirac distribution, which is defined as $N_{\mathrm{FD}}(x) := (\exp(\beta x) + 1)^{-1}$. Since we are considering a bath of molecules at room temperature, we are allowed to take the limit $N_{\mathrm{FD}}(x) \to N_{\mathrm{MB}}(x) := \exp(-\beta x)$. Furthermore, we perform a Wick rotation to obtain the retarded self-energy. The only term that changes in the self-energy is

$$\frac{1}{-i\hbar\omega_n - \epsilon(\mathbf{p}) + \mu_\downarrow + \hbar\omega'} \to i\pi\delta(\hbar\omega' - \epsilon(\mathbf{p}) + \mu_\downarrow - \hbar\omega) + \mathcal{P}\left(\frac{1}{\hbar\omega' - \epsilon(\mathbf{p}) + \mu_\downarrow - \hbar\omega}\right), \tag{27}$$

with the symbol $\mathcal{P}$ symbolizing the principal value of the fraction. With the help of the relationships above, we find for the imaginary part of the self-energy

$$\begin{aligned}
R(\mathbf{k}_+, \omega) &:= -\mathrm{Im}(\hbar\Sigma^{(+)}(\mathbf{k}_+,\omega)) \\
&= -\frac{\pi g_{\mathrm{mol}}^2}{V} \sum_{\mathbf{p}} \int_{-\infty}^{\infty} d\hbar\omega' \rho_\uparrow(\mathbf{p}+\mathbf{k}_+,\omega') \delta(\hbar\omega' - \epsilon(\mathbf{p}) + \mu_\downarrow - \hbar\omega) \left(e^{-\beta\hbar\omega'} - e^{-\beta(\epsilon(\mathbf{p})-\mu_\downarrow)}\right) \\
&= \frac{\sqrt{\pi} g_{\mathrm{mol}}^2 \beta \exp(\beta\mu_\downarrow)(1 - \exp(-\beta\hbar\omega))}{\Lambda^3 \sqrt{2(2\beta\epsilon(\mathbf{k}_+) + (\hbar\beta\Gamma)^2)}} \exp\left(\frac{-\beta(\epsilon(\mathbf{k}_+) + \Delta - \Delta\mu - \hbar\omega)^2}{2((\hbar\beta\Gamma)^2 + 2\beta\epsilon(\mathbf{k}_+))}\right),
\end{aligned} \tag{28}$$

with $\mathbf{k}_+ = (0,0,k_z)$ the wave number for photons in the condensate. By setting $\Delta\mu = 0$ and $k_z = \omega/c_{\mathrm{med}}$ and taking the part of the expression above proportional to $N_\downarrow$, we obtain the absorption cross section at equilibrium. As it

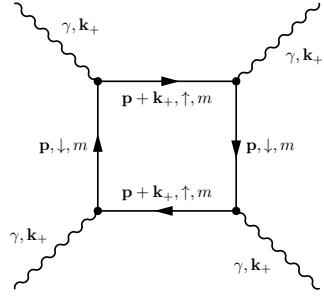

FIG. 4. Feynman diagram for a fourth-order photon-photon interaction. The photons $\gamma$ are considered to be part of the condensate and are thus at zero frequency and momentum $\mathbf{k}_+ = (0,0,k_z)$, as their z-component momentum is fixed and $k_x = k_y = 0$ for the ground state of the homogeneous photon gas. The molecule forms a closed loop of ground ($\downarrow$) and excited ($\uparrow$) states, with momentum $\mathbf{p}$ and Matsubara frequency $\omega_m$.

turns out the result is almost (or in very good approximation) independent of the exact value $k_z$, and therefore from now onwards we take $k_z = 0$, thereby simplifying our treatment considerably. In this case we find

$$\sigma(\omega)\big|_{k_z=0} = \frac{\sqrt{2\pi} g_{\text{mol}}^2}{\hbar^2 \Gamma c_{\text{med}}} \exp\left(-\frac{(\hbar\omega - \Delta)^2}{2\hbar^2 \Gamma^2}\right). \tag{29}$$

Now we compare the absorption cross section for $k_z = 0$ to the experimental data for the fluorescent dye [8, 9]. The absorption spectrum for the dye is asymmetric and impossible to fully reproduce within this simple treatment. However, we can perform a fit to the line shape and obtain the coupling constant $g_{\text{mol}}$, the lifetime $\Gamma$ and detuning $\Delta$. Typical values we find are $g_{\text{mol}} = 4 \cdot 10^{-33}$ J m$^{3/2}$, $\Gamma = 1.15 \cdot 10^{14}$ Hz and $\Delta = 3.8 \cdot 10^{-19}$ J.

### B. Box-Diagram

We now evaluate the photon-photon interacting strength in the condensate, i.e., with Matsubara frequency zero, and set $k_z = 0$, as is justified by the previous section. This amounts to the Feynman diagram in Fig. 4. By using $G_\downarrow^2(\mathbf{p}, m) = -\hbar \partial_\downarrow G_\downarrow(\mathbf{p}, m)$, we obtain

$$\Gamma^{(4)}(\mathbf{0},\mathbf{0},\mathbf{0},0,0,0) = \frac{g_{\text{mol}}^4}{\hbar^3 \beta V} \partial_\downarrow \sum_{\mathbf{p},m} G_\uparrow^2(\mathbf{p},m) G_\downarrow(\mathbf{p},m)$$

$$= \frac{g_{\text{mol}}^4 \partial_\downarrow}{2\pi\beta\hbar^4 \Gamma^2 \Lambda^3} \int_{-\infty}^{\infty} d\hbar\omega \int_{-\infty}^{\infty} d\hbar\omega' \exp\left(\beta\mu_\downarrow - \frac{1}{2(\hbar\Gamma)^2}\left((\hbar\omega + \Delta\mu - \Delta)^2 + (\hbar\omega' + \Delta\mu - \Delta)^2\right)\right)$$

$$\times \frac{1}{\omega - \omega'}\left(\frac{1}{\omega'}(1 - e^{-\beta\hbar\omega'}) - \frac{1}{\omega}(1 - e^{-\beta\hbar\omega})\right) \tag{30}$$

where we again performed the Matsubara summation, we took the limit $N_{\text{FD}}(x) \to N_{\text{MB}}(x)$ and finally performed the $\mathbf{p}$-integral. After we have substituted Eq. (25), we perform the differentiation. Setting $\Delta\mu - \Delta := \mu - \delta$ and introducing the dimensionless quantities $\omega := \beta\hbar\omega$, $\omega' := \beta\hbar\omega'$, $\delta := \beta\delta$ and $\mu := \beta\mu$, we obtain

$$\Gamma^{(4)}(\mathbf{0},\mathbf{0},\mathbf{0},0,0,0) = \frac{g_{\text{mol}}^4 \beta n_{\text{mol}}}{2\pi\hbar^2\Gamma^2\{1 + \exp\left(\mu - \delta + (\beta\hbar\Gamma)^2/2\right)\}} \int_{-\infty}^{\infty} d\omega \int_{-\infty}^{\infty} d\omega' (\omega - \omega')^{-1}$$

$$\times \exp\left(-\frac{1}{2(\beta\hbar\Gamma)^2}\left((\omega + \mu - \delta)^2 + (\omega' + \mu - \delta)^2\right)\right)$$

$$\times \left(\frac{1}{\omega'}(1 - e^{-\omega'}) - \frac{1}{\omega}(1 - e^{-\omega})\right)\left(1 + \frac{1}{(\beta\hbar\Gamma)^2}(\omega + \omega' + 2\mu - 2\delta)\right)$$

$$:= \frac{g_{\text{mol}}^4 \beta n_{\text{mol}}}{\hbar^2 \Gamma^2} f(\mu - \delta), \tag{31}$$

with $f(\mu - \delta)$ a smooth dimensionless function peaked around zero. Again we must scale $\Gamma^{(4)} \to 2\Gamma^{(4)}/3D_0$, as the photon gas is confined to two dimensions, to obtain the effective coupling constant $g$ of the photons in the condensate.



\begin{thebibliography}{99}
\bibitem{Beenakker}
	C. Beenakker and C. Sch\"onenberger,
	Phys. Today $\vec{56}$, 37 (2003).

\bibitem{Schottky}
	W. Schottky, 
	Ann. Phys. $\vec{57}$, 541 (1918).

\bibitem{Cooper}
	F. Lefloch, C. Hoffmann, M. Sanquer and D. Quirion,
	Phys. Rev. Lett. $\vec{90}$, 067002 (2003). 

\bibitem{Frac}
	R. de Picciotto et al.,
	Nature $\vec{389}$, 162 (1997).

\bibitem{Nyquist}
	H. Nyquist,
	Phys. Rev. $\vec{32}$, 110 (1928).

\bibitem{Johnson}
	J. Johnson, 
	Phys. Rev. $\vec{32}$, 97 (1928).

\bibitem{CalWel}
	H.B. Callen and T.A. Welton
    Phys. Rev. $\vec{83}$, 34 (1951).

\bibitem{Kubo}
	R. Kubo, Rep. Prog. Phys. $\vec{29}$, 255 (1966). 

\bibitem{AtomicBEC1}
	M. H. Anderson, J.R. Ensher, M.R. Matthews, C.E. Wieman, E. A. Cornell,
	Science $\vec{269}$, 5221 (1995).

\bibitem{Noise}
	E. Altman, E. Demler and M. Lukin,
	Phys. Rev. A $\vec{70}$, 013603 (2004).

\bibitem{AtomicBEC2}
	N. Cherroret and S.E. Skipetrov,
	Phys. Rev. Lett. $\vec{101}$, 190406 (2008).

\bibitem{QuasiBEC1}
	J. Kasprzak et al.,
	Nature $\vec{443}$, 409 (2006). 

\bibitem{QuasiBEC2}
	S. O. Demokritov et al.,
	Nature $\vec{443}$, 430 (2006).

\bibitem{Weitz}
	J. Klaers, J. Schmitt, F. Vewinger and M. Weitz, 
	Nature $\vec{468}$, 545 (2010).

\bibitem{Weitz3}
	J. Klaers, J. Schmitt, T. Damm, F. Vewinger and M. Weitz,
	Appl. Phys. B $\vec{105}$, 17 (2011).

\bibitem{Weitz2}
	J. Schmitt, T. Damm, D. Dung, F. Vewinger, J. Klaers and M. Weitz, 
	Phys. Rev. Lett. $\vec{112}$, 030401 (2014).

\bibitem{Stoof}		
 	H.T.C. Stoof, K.B. Gubbels and D.B.M. Dickerscheid, 
 	\emph{Ultracold Quantum Fields}, 
 	Springer (2009).

\bibitem{GaussianAtomic}
 	G. Baym and C. J. Pethick,
	Phys. Rev. Lett. $\vec{76}$, 6 (1996).

\bibitem{Smith}
	H. Smith and C.J. Pethick,
	\emph{Bose-Einstein Condensation in Dilute Gases},
	Cambridge University Press, Cambridge,
	2nd Edition (2008).

\bibitem{Private}
	Private correspondence with J. Klaers.

\bibitem{Engeland}
	R. A. Nyman and M. H. Szymanska,
	arXiv:1308.3588 [quant-ph] (2013).

\bibitem{HandBook}
	\emph{Handbook of Chemistry and Physics},
	CRC Press, 91st Edition (2009).

\bibitem{Turkish}
	S. Yaltkaya and R. Aydin,
	Turk. J. Phys. $\vec{22}$, 41 (2002).

\bibitem{PhotonPhoton}
	R.Y. Chiao, T.H. Hansson, J.M. Leinaas and S. Viefers,
	Phys. Rev. A. $\vec{69}$, 063816 (2004).

\bibitem{AW}
	A.-W. de Leeuw, H.T.C. Stoof and R.A. Duine, 
	Phys. Rev. A $\vec{88}$, 033829 (2013).

\bibitem{Feshbach}
	C. Chin, R. Grimm, P. Julienne and E. Tiesinga,
	Rev. Mod. Phys. $\vec{82}$, 1225 (2010).

\end{thebibliography}
\end{document}